\documentclass[3p,times,procedia]{elsarticle}
\usepackage{nupha_ecrc}

\volume{00}

\firstpage{1}

\journalname{Nuclear Physics A}

\runauth{}

\jid{nupha}

\jnltitlelogo{Nuclear Physics A}

\usepackage{amssymb}

\usepackage[figuresright]{rotating}

\begin{document}

\begin{frontmatter}



\dochead{XXVIIth International Conference on Ultrarelativistic Nucleus-Nucleus Collisions\\ (Quark Matter 2018)}

\title{Applications of deep learning to relativistic hydrodynamics}


\author[label1,label2]{Hengfeng Huang}
\author[label3]{Bowen Xiao}
\author[label1]{Huixin Xiong}
\author[label1,label2]{Zeming Wu}
\author[label3,label4]{Yadong Mu}
\author[label1,label2,label5]{Huichao Song}
\address[label1]{Department of Physics and State Key Laboratory of Nuclear Physics and Technology, Peking University, Beijing 100871, China}
\address[label2]{Collaborative Innovation Center of Quantum Matter, Beijing 100871, China}
\address[label3]{Institute of Computer Science and Technology, Peking University, Beijing 100080, China}
\address[label4]{{Center for Data Science}, Peking University, Beijing 100871, China}
\address[label5]{Center for High Energy Physics, Peking University, Beijing 100871, China}

\begin{abstract}
In this proceeding, we will briefly review our recent progress on implementing deep learning to relativistic hydrodynamics. We will demonstrate that a successfully designed and trained deep neural network, called {\tt stacked U-net}, can capture the main features of the non-linear evolution of hydrodynamics, which could also
rapidly predict the final profiles for various testing initial conditions.
\end{abstract}


\end{frontmatter}


\section{Introduction}
Deep learning is one of the machine learning technic based on the deep neural network. Recently, it has been successfully applied to many research areas in physics, such as the search of gravitational lens~\cite{Hezaveh:2017sht}, identifying and classifying the phases of Ising model~\cite{Carrasquilla:2017}, the search of Higgs and exotic particles~\cite{Baldi:2014kfa}, classification of jet structure~\cite{Komiske:2016rsd}, identifying the equation of state of hot QCD matter~\cite{Pang:2018}, etc.

In this proceeding, we will introduce our recent work on implementing deep learning to relativistic hydrodynamics~\cite{Huang:2018fzn}.  Relativistic hydrodynamics is a powerful tool to simulate the evolution of the quark-gluon plasma (QGP) and to study various flow observables in relativistic heavy ion collisions~\cite{Heinz:2013th}. In the traditional method, hydrodynamic simulations numerically solve the non-linear transport equations from the conservation laws and the second law of thermodynamics, which translate the initial conditions to final profiles through the non-linear evolution~\cite{Heinz:2013th,Song:2007ux}. In our recent work~\cite{Huang:2018fzn}, we designed and trained a deep neural network, called {\tt stacked U-net}, to learn the non-linear mapping between initial and final profiles from hydrodynamic simulations with {\tt MC-Glauber} initial conditions, which can also rapidly predict the final profiles for different initial conditions, including {\tt MC-KLN}, {\tt AMPT} and {\tt TRENTo}.

\begin{figure}[t]
	\centerline{\includegraphics[width=1.0\linewidth]{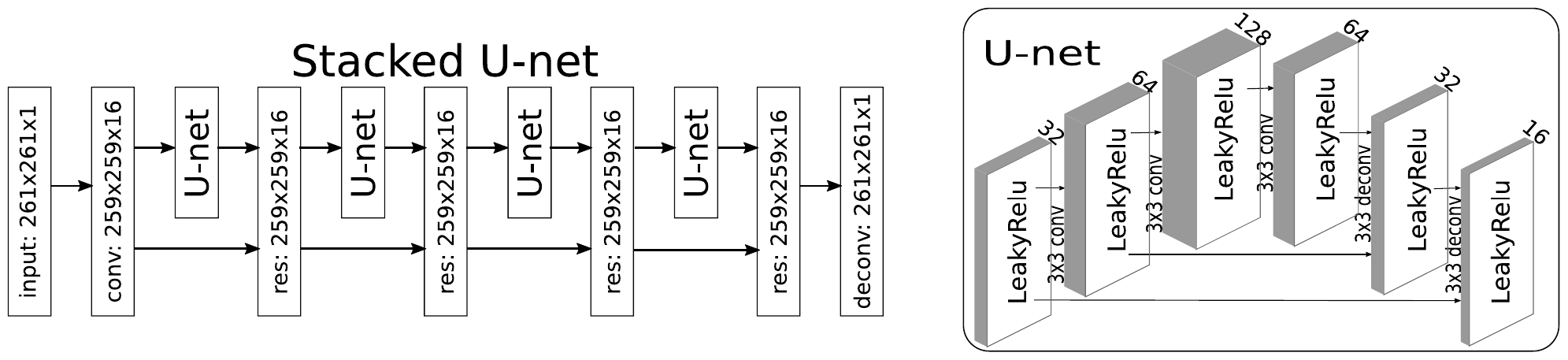}}
	\caption{The architecture of {\tt stacked U-net} that consists of four residual U-net blocks. The right panel shows the U-net structure, and the depth of the hidden layer is denoted on the top of them. }
	\label{net}
\end{figure}

\section{Method}
The training and testing data sets for deep learning are generated from traditional hydrodynamic simulations. For simplicity, we implemented (2+1)-d ideal hydrodynamics with zero viscosity and charge densities, using the code {\tt VISH2+1} with an ideal equation of state EoS $p=\frac{e}{3}$~\cite{Song:2007ux}. The initial energy density profiles are generated by some initial condition models, such as {\tt MC-Glauber}~\cite{Miller:2007ri,Hirano:2009ah}, {\tt MC-KLN}~\cite{Hirano:2009ah,Drescher:2006ca}, {\tt AMPT}~\cite{Pang:2012he} and {\tt TRENTo}~\cite{Moreland:2014oya} with zero transverse flow velocity. We run {\tt VISH2+1} at three fixed evolution times $\tau-\tau_0=2.0, \ 4.0, \ 6.0 \ \mathrm{fm/c}\ (\tau_0=0.6\mathrm{fm/c})$ to obtain the final energy momentum tensor profiles $T^{\tau\tau}(\tau,x,y), T^{\tau x}(\tau,x,y),T^{\tau y}(\tau,x,y)$. The time step and grid size of the simulations are set to $d\tau=0.04\ \mathrm{fm/c}$ and $dx=dy=0.1\ \mathrm{fm}$ within a fixed transverse area of $13\ \mathrm{fm}\times 13\ \mathrm{fm}$.

The designed neural network, called {\tt stacked U-net(sU-net)}, belongs to the encoder-decoder network architecture, which could enhance the gradient flow in the deep part of the network during back propagation.
As shown in Fig.~1, it contains four U-net blocks in series with residual connection. Inside each U-net block, there are three convolutional and deconvolutional layers, and the outputs of the first two convolutional layers are also fed into the last two deconvolutional ones through concatenating the feature maps along the channel dimension, respectively. The activation functions for all layers, except for the output one, are \emph{Leaky ReLU} $f\left( x \right) = \max \left\{ {x,0.03x} \right\}$. While the ones for the output layer are \emph{softplus} $f\left( x \right) = \ln \left( {1 + {e^x}} \right)$ for $T^{\tau\tau}$ and $f(x)=1$ for $T^{\tau x}$ and $T^{\tau y}$. The kernel size of all layers is $3\times 3$, and the loss function is \emph{normalized MSE loss}, $Loss = \frac{\| y_1 - y_0 \|^2}{\| y_0 \|^2}$, where $y_1$ is the output of the network and $y_0$ is the ground truth. We minimize the loss function with the standard mini-batch stochastic gradient descent algorithm with batch size selected as 16 and a learning rate exponentially decaying from $10^{-3}$ to $10^{-5}$.

In practice, we divide the whole evolution time $\tau-\tau_0=6.0\ \mathrm{fm/c}$ into three parts with equal time interval $\Delta\tau=2.0\mathrm{fm/c}$, and train three individual {\tt sU-net} for each part. After that, we use the trained {\tt sU-net-1} to predict the profiles at $\tau-\tau_0=2.0\mathrm{fm/c}$, and use them as initial conditions for {\tt sU-net-2} to predict the profiles at $\tau-\tau_0=4.0\mathrm{fm/c}$, and then use {\tt sU-net-3} to predict the final profiles at $\tau-\tau_0=6.0 \ \mathrm{fm/c}$ (please refer to~\cite{Huang:2018fzn} for details).

\begin{figure}[t]
	\centerline{\includegraphics[width=1.0\linewidth]{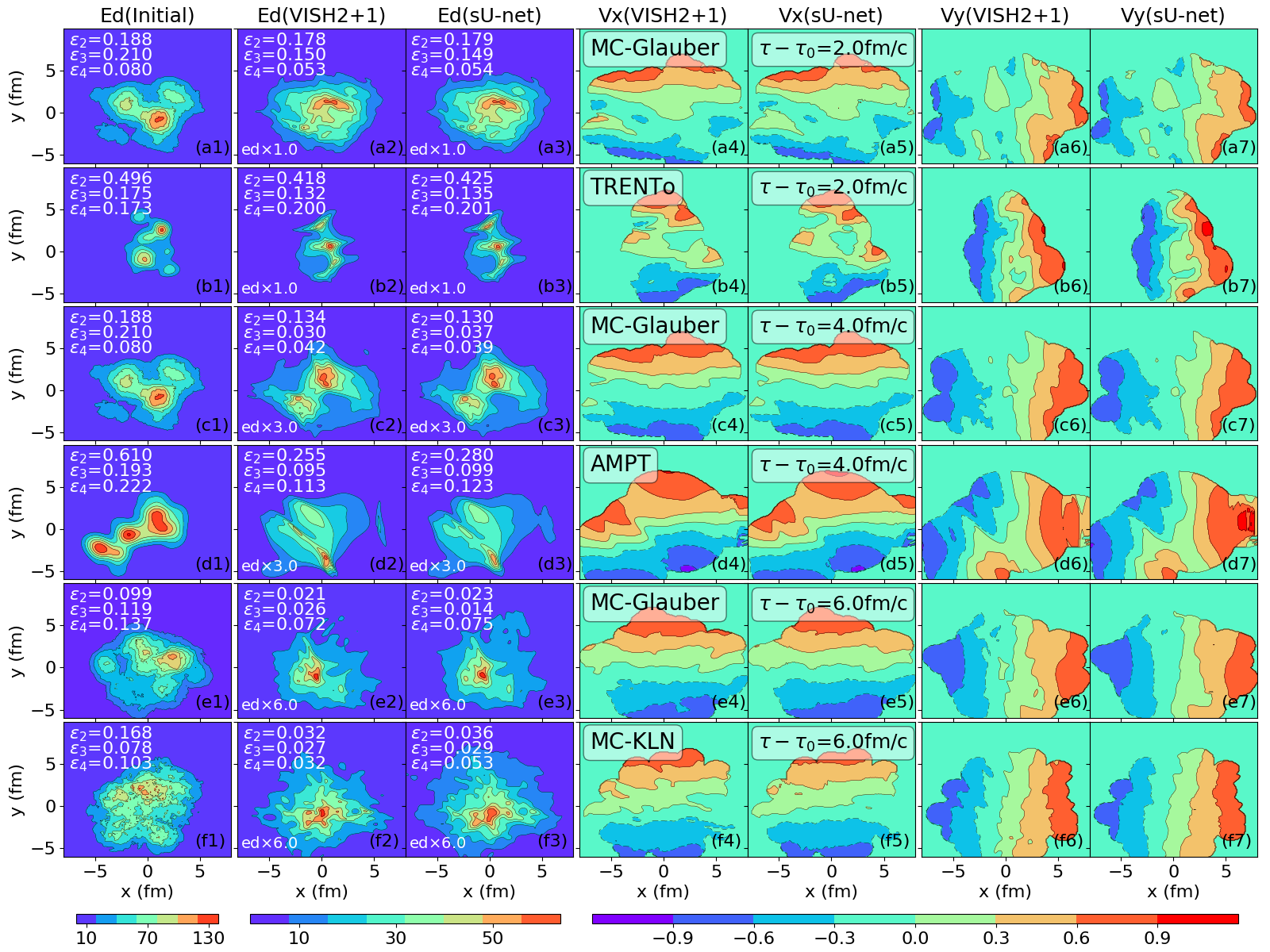}}
	\caption{ Energy density and flow velocity profiles at $\tau-\tau_0=2.0, \ 4.0, \ 6.0 \ \mathrm{fm/c}$, calculated from {\tt VISH2+1} and predicted by {\tt Stacked U-Net} for six test cases with initial profiles generated from {\tt MC-Glauber}, {\tt MC-KLN}, {\tt AMPT} and {\tt TRENTo}.}
	\label{cmp}
\end{figure}

\begin{figure}[t]
	\centerline{\includegraphics[width=0.95\linewidth]{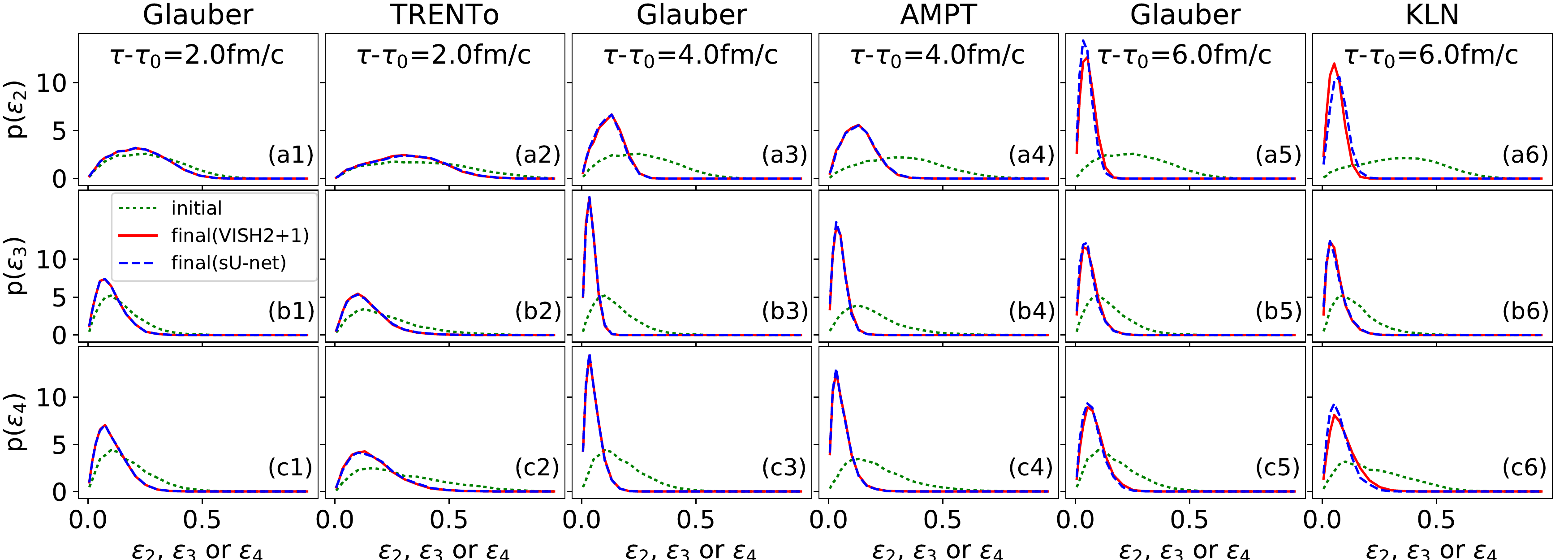}}
	\caption{ Eccentricity distribution $P(\varepsilon_n)$ (n=2, 3, 4), at $\tau-\tau_0=2.0, \ 4.0, \ 6.0 \ \mathrm{fm/c}$, calculated from {\tt VISH2+1} and predicted by {\tt Stacked U-Net} for 100000 tested initial profiles generated from {\tt MC-Glauber}, {\tt MC-KLN}, {\tt AMPT} and {\tt TRENTo}.}
	\label{eccentricity}
\end{figure}

\section{Results}
The network {\tt sU-net} is trained with  10000 initial and final profiles from {\tt VISH2+1} simulations with {\tt MC-Glauber} initial conditions, and then tested on making predictions for the final profiles of four different types of initial conditions, including {\tt MC-Glauber}, {\tt MC-KLN}, {\tt AMPT} and {\tt TRENTo}.  Fig.~2 shows the comparison
between hydrodynamic results and the network predictions on final energy density and flow velocity at $\tau-\tau_0=2.0, \ 4.0, \ 6.0 \ \mathrm{fm/c}$. It shows that the trained network is able to capture the magnitudes and structures of the contour plots of both energy density and flow velocity. Meanwhile, panel (b), (d) and (f) also show that {\tt sU-net}, trained with data sets associated with the {\tt MC-Glauber} initial conditions, can also nicely predict the final profiles of other initial conditions.

We also calculate the the eccentricity coefficients for the energy density profiles, $ \varepsilon_n= \frac{\int r^{n} dr d\phi e(r,\phi) e^{in\phi}}{\int r^{n} dr d\phi e(r,\phi)} \ \ \ (n=2, 3, 4) $, which helps to evaluate the deformation and inhomogeneity of the QGP fireball. These true and predicted values of  $ \varepsilon_n$ (n=2, 3, 4) from hydrodynamics and network are written in the related panels(a-f).

Fig.~3 presents the eccentricity distributions $P(\varepsilon_n)$ of energy density at $\tau-\tau_0=2.0, \ 4.0,  \ 6.0  \ \mathrm{fm/c}$, which shows that the results from the network almost overlap with the ones from {\tt VISH2+1} and are also obviously deviated from the initial eccentricity distributions $P_0(\varepsilon_n)$.  To further evaluate the accuracy of the network predictions, we plot the histograms of the true vs. predicted eccentricity $\varepsilon_n$ at $\tau-\tau_0=2.0, \ 4.0, \ 6.0 \ \mathrm{fm/c}$, calculated from {\tt VISH2+1} and predicted by {\tt stacked U-Net} for 100000 tested initial profiles generated from {\tt MC-Glauber}, {\tt MC-KLN}, {\tt AMPT} and {\tt TRENTo}. The prediction accuracy is very good at shorter evolution time $\tau-\tau_0=2.0\ \mathrm{fm/c}$, but  gradually decreases for longer evolution time.  Due to the limited memory of our local GPU server, we implement the combined {\tt sU-net} series to predict the final profiles at longer evolution time, which also gradually accumulates errors with more {\tt sU-net} added.

\begin{figure}[t]
	\centerline{\includegraphics[width=0.95\linewidth]{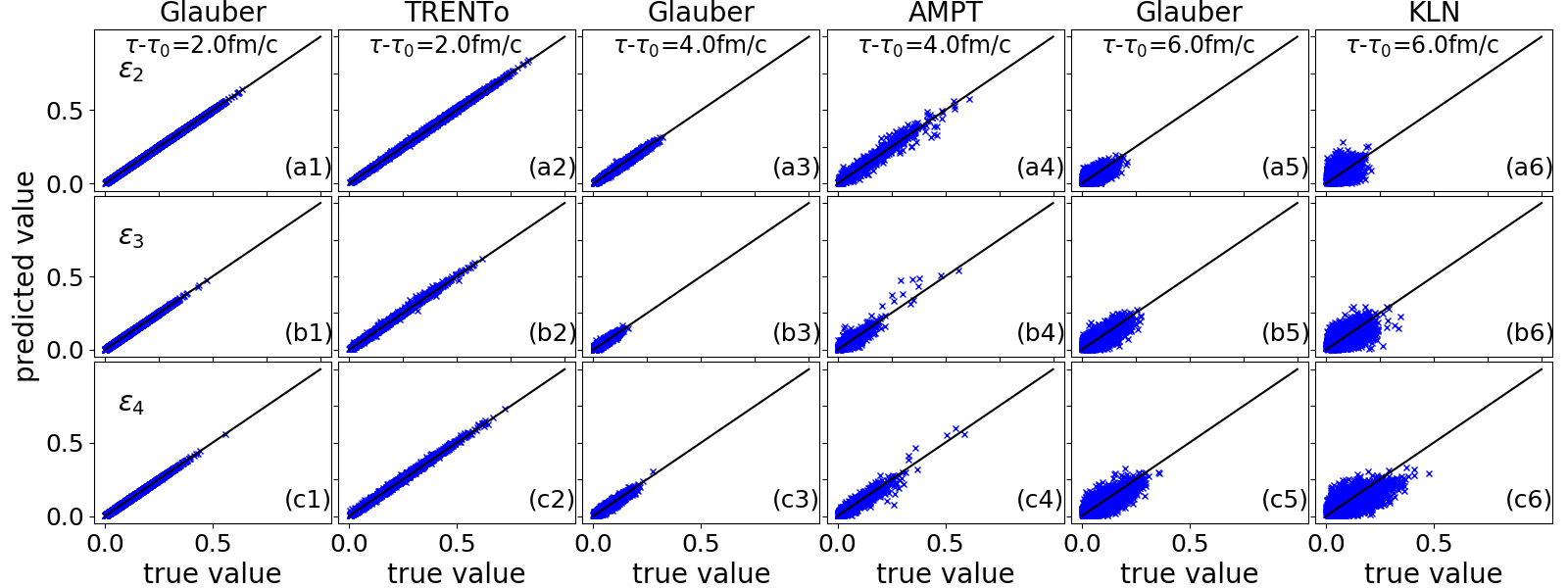}}
	\caption{ Histograms of the true vs. predicted eccentricity $\varepsilon_n$ at $\tau-\tau_0=2.0, \ 4.0, \ 6.0 \ \mathrm{fm/c}$, calculated from {\tt VISH2+1} and predicted by {\tt Stacked U-Net} for 100000 tested initial profiles generated from {\tt MC-Glauber}, {\tt MC-KLN}, {\tt AMPT} and {\tt TRENTo}.}
	\label{ecce_diff}
\end{figure}

\section{Conclusions}
In this proceeding, we demonstrate that the well designed and trained neural network can capture the main features of the non-linear evolution of the relativistic hydrodynamics, which can not only predict the magnitude and inhomogeneous structures of the final profiles, but also describe the related eccentricity distributions $P(\varepsilon_n)$ (n=2, 3, 4). We also noticed that prediction accuracy decreases with longer evolution time, which should be further improved with new designed network structure and more powerful GPU server.

\section{Acknowledgments}
We thank the discussions from  L.G. Pang and K. Zhou. H.~H, H.~X. Z.~W. and H.~S. are supported by the NSFC and the MOST under grant Nos.11435001, 11675004 and 2015CB856900. B.~X. and Y.~M. are supported by grant no.~61772037.

\bibliographystyle{elsarticle-num}

\end{document}